\documentclass[twocolumn,amsmath,amssymb,groupedaddress]{revtex4}

\usepackage{graphicx}
\usepackage{amssymb}
\usepackage{epstopdf}
\usepackage{bbold}
\usepackage{xcolor}

\usepackage{soul}

\usepackage{hyperref}

\hypersetup{%
pdftitle=\@title,
citecolor=Blue,
filecolor=black,
linkcolor=Blue,
urlcolor=Blue,
breaklinks,
hidelinks,
bookmarksopen=true,
}%

\newcommand*{\doi}[1]{\sloppy\href{http://dx.doi.org/#1}{doi:~#1}}

\begin{document}
\title{Elasto-plastic description of brittle failure in amorphous materials}

\author{Marko Popovi{\'c}}
\author{Tom~W.~J.\ de Geus}
\author{Matthieu Wyart}
\address{Institute of Theoretical Physics, \'{E}cole Polytechnique F\'{e}d\'{e}rale de Lausanne (EPFL), CH-1015 Lausanne, Switzerland}

\begin{abstract}
The response of amorphous materials to an applied strain can be continuous, or instead display a macroscopic stress drop when a shear band nucleates. Such discontinuous response can be observed if the initial configuration is very stable. We study theoretically how such brittleness emerges in athermal, quasi-statically driven, materials as their initial stability is increased. We show that this emergence is well reproduced by elasto-plastic models and is predicted by a mean field approximation, where it corresponds to a continuous transition. In mean field, failure can be forecasted from the avalanche statistics. We show that this is not the case for very brittle materials in finite dimensions due to rare weak regions where a shear band nucleates. Their critical radius is predicted to follow $a_c\sim (\Sigma-\Sigma_b)^{-2}$, where $\Sigma$ is the stress and $\Sigma_b$ the stress a shear band can carry. 

\end{abstract}

\maketitle


How amorphous solids such as granular materials, bulk metallic glasses, colloidal suspensions or foams yield under an applied strain is a central question in fields as diverse as geophysics \cite{Johnson05}, material science \cite{Schroers04} and soft matter \cite{Bonn15}. At a macroscopic level, the stress {\it vs} strain curve under quasi-static loading can monotonically increase or slightly overshoot as in foams and granular materials \cite{Andreotti13} or even be discontinuous as in some metallic glasses \cite{Antonaglia14}. This brittleness can have a catastrophic consequence, and appears to depend on a variety of factors including composition \cite{Lewandowski05}, Poisson's ratio \cite{Lewandowski05}, temperature \cite{Lu03} and system preparation \cite{Gu09}. Spatially, it corresponds to the emergence of a shear band a few nanometers thick \cite{Zhang06} where most of the strain localizes. Understanding what aspects of the material ultimately control brittleness and how shear bands nucleate remains a challenge. At a microscopic level, plasticity takes place by discrete events, the so-called shear transformations where a few particles rearrange locally \cite{Argon79,Falk98,Schall07,Nicolas17}. The stress change it induces is anisotropic and long-range \cite{Picard04,Nicolas17}, and can in turn trigger new plastic events, generating anisotropic avalanches of plasticity \cite{Lemaitre09,Maloney09}. It has been argued that amorphous solids are critical: plasticity in the solid phase occurs via avalanches that can be system spanning \cite{Lin15,Muller14,Budrikis17}. Yet it is unclear if these avalanches of plasticity are precursors of brittle failure \cite{Gimbert12, Le-Bouil14}.

Recently, novel algorithms have been able to generate very stable glasses that are brittle. This previously impossible feat was achieved by obtaining quench rates comparable to experiments \cite{Ozawa18,Jin18} using swap algorithms \cite{Ninarello17,Brito18}, or by shearing the system back and forth many times \cite{Leishangthem17}. These studies underline the critical role of the system preparation in controlling brittleness. Theoretically, it was recently proposed that the yielding transition is a spinodal decomposition \cite{Rainone15, Parisi17}, which occurs for example in a magnet if a field is applied in the direction opposed to its magnetization. The magnetization can evolve smoothly (``ductile'' behavior) or rather suddenly (``brittle'' behavior) depending on the amount of disorder \cite{Sethna93,Nandi16}. This analogy nicely explains why increasing the initial stability of the glass can lead to a ductile to brittle transition \cite{Ozawa18}. Yet, it does not incorporate the anisotropy of the interaction between shear transformations that causes shear bands, nor the criticality of the solid phase.

In this Letter, we first show that this ductile to brittle transition is well captured by elasto-plastic models \cite{Nicolas17} by increasing the stability of the initial configurations. We explain this observation in a mean field approximation where the transition is found to be continuous, and failure can indeed be anticipated from the distribution of avalanches. We then argue that these results break down for very stable glasses due to rare locations in the sample where a shear band nucleates: failure occurs if the spatial extension $a$ of a weak region in the sample exceeds $a_c\sim (\Sigma-\Sigma_b)^{-2}$, where $\Sigma$ is the stress and $\Sigma_b$ is the stress that a shear band can carry. We confirm these results in elasto-plastic models, both by measuring the effect of inserting a weak ``scar'' in the material and by studying finite size effects. Overall, the framework we propose for how amorphous solids yield in quasi-static athermal conditions ties together ductile and brittle behavior, avalanche statistics and shear band nucleation of very brittle materials.

\paragraph*{Brittleness in elasto-plastic models}
In elasto-plastic models \cite{Baret02, Picard05,Nicolas17} the material is divided into $N$ elements, each characterized by its shear stress $\sigma_i$ and yield stress  $\sigma_i^Y$. The overall stress of the system is simply $\Sigma= \sum_i\sigma_i/N$. When $|\sigma_i|$ reaches $\sigma_{i}^Y$, the element yields: after a time $\tau$ its stress decreases by a value $\delta \sigma_i$, corresponding to a plastic deformation $\delta \epsilon_{p, i}= \delta\sigma_i/\mu_0$, where $\mu_0$ is the shear modulus. New random variables $\sigma_i$ and $\sigma^Y_i$ are then taken from some distributions $P(\sigma)$ and $P_Y(\sigma^Y)$. Such a plastic event affects the stress everywhere in the material, according to a propagator $G(\vec{r})$ whose sign varies in space and which decays in magnitude as a dipole \cite{Picard05,Nicolas17,Demery17}. The specific parameters we use are described in the Supplementary Material. Such models have a finite macroscopic yield stress $\Sigma_c$ so that the material is solid for $|\Sigma| < \Sigma_c$ and liquid for $|\Sigma| > \Sigma_c$ \cite{Nicolas17}. In the solid case these models predict how $\Sigma$ depends on the accumulated plastic strain $\epsilon_p= \sum_i \epsilon_{p,i}/N$.

As the stress $\Sigma$ is increased, most elements yield by reaching $\sigma_i= + \sigma_i^Y$. Therefore, it is useful to characterize elements by their stability $x_i = \sigma_i^Y-\sigma_i$. Depending on the initial distribution of stability $P_0(x)$, the stress was found to overshoot or not \cite{Jagla07, Lin15}. However, a brittle behavior has not been reported within these models.

\begin{figure}[ht!]
\centering
\includegraphics[width=.48\textwidth]{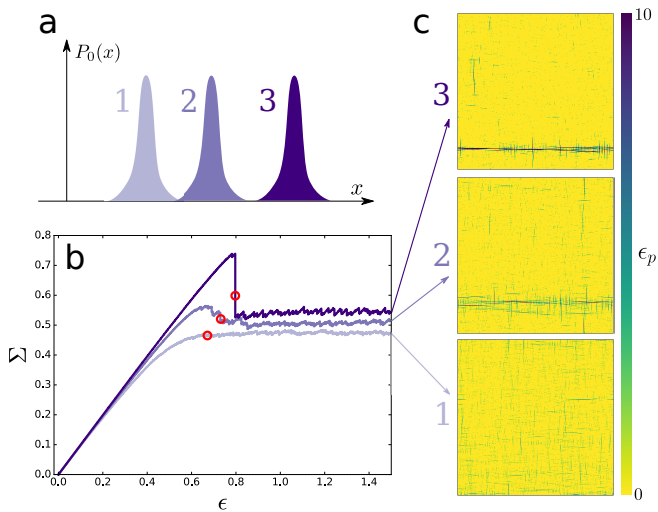}
\caption{An elasto-plastic model can be ductile without ($1$) and with ($2$) a stress overshoot, or brittle ($3$), depending on the preparation. a) Schematic of the initial distribution of stabilities, $P_0(x)$, representing the three system preparations. b) Stress strains curves for these three different $P_0(x)$ together with snapshots of the spatial distribution of plastic strain. c) 1: If the system is preparation is not very stable it will shear homogeneously and there will be no stress overshoot. 2: As the stability of preparation is increased a shear band is formed. 3: For a very stable preparation, a sharp shear band is formed during the brittle failure. Snapshots of the plastic strain are taken at positions indicated by red circles on stress {\it vs} strain curve.}
\label{fig:spatialResults}
\end{figure}

We proceed by increasing stability of the distribution $P_0(x)$ as illustrated in Fig.\ \ref{fig:spatialResults}a. For weakly stable initial states (case $1$), the strain is homogeneous and the stress does not overshoot. When the initial stability is increased (case $2$ in Fig.\ \ref{fig:spatialResults}) the stress {\it vs} strain curve does display an overshoot. Although there is no macroscopic drop of stress, avalanches tend to localize along a rather thin shear band, as justified in \cite{Manning07,Moorcroft13}.

A key observation is that for very stable systems, the scenario changes: the stress {\it vs} strain curve becomes discontinuous (case $3$ in Fig.\ \ref{fig:spatialResults}). A very narrow shear band appears in one single avalanche, and relaxes the stress by some finite amount which persists in the thermodynamic limit, see below. This result supports that macroscopic failure can occur even in the absence of inertia and thermal feedback (in which strain increases temperature locally, which in turn localizes strain further), as these effects are absent in our model.

\begin{figure}[ht!]
\centering
 \includegraphics[width=.48\textwidth]{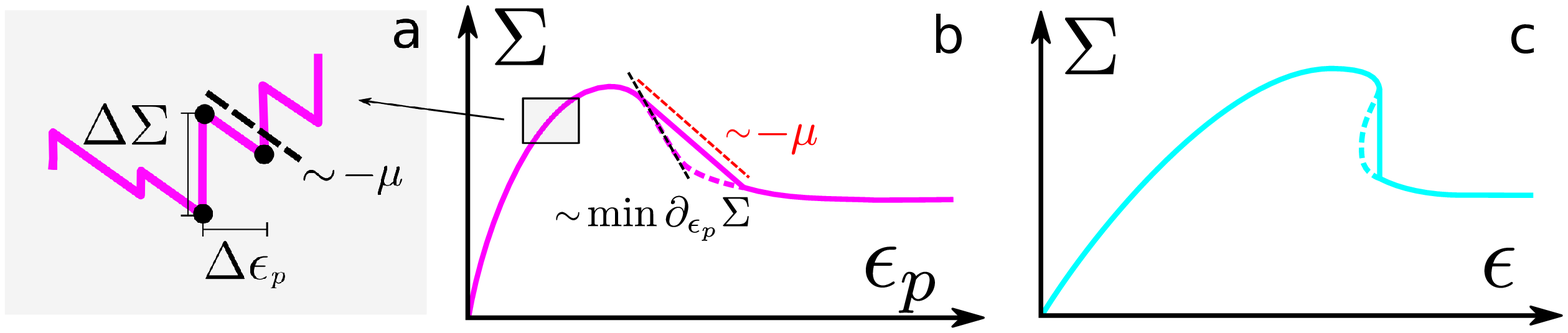}
 \caption{a) Stress {\it vs} plastic strain curve consists of alternating elastic stress increases $\Delta\Sigma$ and stress relaxations by an avalanche of plastic events $\Delta\Sigma_{\mathrm{avalanche}}= -\mu \Delta\epsilon_p$. b) When the slope of the macroscopic stress {\it vs} plastic strain curve reaches $-\mu$ an extensive avalanche occurs. c) Stress {\it vs} strain curve corresponding to the stress {\it vs} plastic strain curve from b) during a strain controlled loading.}
\label{fig:generalConsiderations}
\end{figure}

\paragraph*{Avalanches and macroscopic failure}
To explain this finding, we first consider the relationship between the avalanche size $S \equiv N \Delta \epsilon_p$, where $\Delta \epsilon_p$ is the total plastic strain accumulated during the avalanche, and the stress {\it vs} strain curve. When elements in the system begin to fail and the system deforms plastically, $P(x)$ develops a pseudo-gap $P(x)\sim x^\theta$ with $\theta>0$ \cite{Lin14a,Lin16}. This result implies in turn that the minimal stability in the entire system (characterizing the size of the elastic ramps in Fig.\ \ref{fig:generalConsiderations}a) follows $x_{min}\sim N^{-1/(1+\theta)}$ \cite{Karmakar10a}, which was shown to constrain avalanche statistics for stress-controlled loading \cite{Lin15}.

We generalize this result by noting that controlling stress is a special case of a more general loading protocol in which a spring of elasticity $\mu_S$ is placed between the system and a strain controlled loading apparatus. Stress controlled loading then corresponds to $\mu_S \rightarrow 0$ and strain controlled loading to $\mu_S\rightarrow \infty$. The overall shear elastic constant of this combined system is $\mu= \mu_0 \mu_S/(\mu_0 + \mu_S)$, equivalent to a serial connection of two springs with elastic constants $\mu_0$ and $\mu_S$. In practice, $\mu_S$ has a finite value, and experimental protocols lie between these limits. Consider an increment of stress $\Delta \Sigma=x_{min}$ followed by an avalanche where the stress drops by $\Delta\Sigma_{\mathrm{avalanche}} = -\mu \Delta \epsilon_p$, which appears as a kink highlighted by the three black points in Fig.\ \ref{fig:generalConsiderations}a. Requiring that on average this kink has an overall slope $\partial\Sigma/\partial \epsilon_p$, and using the definition of $S$ as well as the scaling for $x_{min}$ it follows that
\begin{align}
\label{eq:averageAvalanche}
  \langle S\rangle \sim\frac{N^{\frac{\theta}{\theta + 1}}}{1+ \frac{1}{\mu}\frac{\partial \Sigma}{\partial \epsilon_p} } \quad .
\end{align}
Key consequences of Eq.\ \eqref{eq:averageAvalanche} are (i) $\partial \Sigma/\partial \epsilon_p=-\mu$ is a sufficient condition for macroscopic failure (not always necessary, see below). Thus, if the spring $\mu_S$ is stiff, macroscopic failure is less likely: in particular if $\min_{\epsilon_p} \partial \Sigma/\partial \epsilon_p>-\mu$ we predict no macroscopic failure. (ii) The mean avalanche size generically diverges with $N$, signaling crackling noise and system spanning avalanches even away from failure. This result is qualitatively different from disordered magnets \cite{Sethna93} where the approach of failure is required for crackling to occur. However, the denominator in \eqref{eq:averageAvalanche} diverges as the criterion $\partial \Sigma/\partial \epsilon_p\rightarrow -\mu$ is approached, suggesting that failure may be anticipated by monitoring avalanches.

Henceforth we shall focus on the strain controlled set-up, where Eq.\ \eqref{eq:averageAvalanche} becomes $\langle S\rangle \sim N^{\frac{\theta}{\theta + 1}} (1-(\partial \Sigma/\partial \epsilon)/\mu)$ where $\epsilon$ is the total strain $d\epsilon=d\epsilon_p +d\Sigma/\mu$. Then a sufficient condition for failure is that the stress {\it vs} strain curve develops an infinite slope, as illustrated in Fig.\ \ref{fig:generalConsiderations}c. Interestingly, it is still possible to probe this curve when it overhangs, if we allow the set-up to have a negative stiffness $\mu_S< 0$, as is the case in the formalism we now develop.

\begin{figure}[ht!]
\centering
 \includegraphics[width=.48\textwidth]{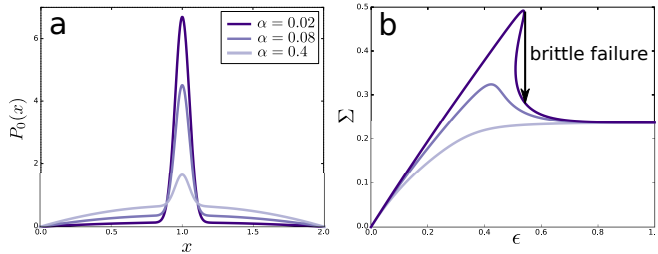}
\caption{a) Initial stability distributions $P_0(x) \sim (1 - \alpha)\exp(-(x-1)^2/(2s_P^2)) + \alpha x (2 - x)$ with $s_P= 0.05$ we use to find b) stress {\it vs} strain curves in the Hebraud-Lequeux model, showing the ductile to brittle transition as the initial stability distribution $P_0(x)$ is narrowed.}
\label{fig:meanField}
\end{figure}

\paragraph*{Mean field approximation} Following the previous paragraph, brittleness can be predicted by computing $\partial \Sigma/\partial \epsilon_p$.
This is very hard in general because the mechanical noise generated by shear transformations is highly correlated in space. Mean field approximations neglect these correlations \cite{Hebraud98}. In its simplest form, the mechanical noise is assumed to be white, corresponding to the Hebraud-Lequeux model \cite{Hebraud98}. In more realistic mean field models, the noise is much broader, which leads to better values for the pseudo-gap exponent $\theta$ \cite{Lin16}. For our present purpose, however, we expect, and have checked numerically, that the two models lead to qualitatively similar behavior. We thus consider the simpler Hebraud-Lequeux model.

For simplicity, we assume yield stresses to be identical for all elements, and set its value $\sigma^Y$ to unity. We further assume that locally the material is fully plastic, so that $\sigma_i \rightarrow 0$ and $x_i\rightarrow 1$ once element $i$ yields. Thus $x_i=0$ and $x_i=2$ corresponds to the limit of stability of elements and elements that have yielded are reintroduced at $x_i=1$. With this notation, the total stress is $\Sigma= 1 - \int_0^2 x P(x) dx$. The dynamical equation for the stability distribution $P(x)$ is a diffusion equation \cite{Hebraud98}
\begin{align}
\label{eq:HBquasistatic}
\partial_{\gamma}P(x, \gamma)&= D \partial_x^2P(x, \gamma) + v \partial_xP(x, \gamma) + \delta(x-1) \ \  .
\end{align}
Here, $\gamma\equiv \epsilon_p\mu/\sigma^Y$ is number of plastic events per element, $D$ characterizes the amplitude of the mechanical noise, and the source term describes the reinsertion of elements that have yielded. The drift $v$ is a Lagrange multiplier that allows us to impose quasi-static loading.
It is prescribed as follows: during quasi-static loading no elements are unstable in the thermodynamic limit, implying the boundary conditions $P(0)= P(2)= 0$.
This condition precludes failure, which will instead be signaled by an overhanging stress {\it vs} strain curve.
By integrating Eq.\ \eqref{eq:HBquasistatic} we find that $\partial_xP(2, \gamma) - \partial_xP(0, \gamma) = -1/D$. In practice, the first term becomes very small as soon as the stress rises \cite{Lin16} because almost no sites yield in the ``wrong'' direction at $x=2$. Therefore, we can neglect the first term so that $\partial_xP(0, \gamma) = 1/D$. Taking the derivative of the stress, we now find
\begin{align}
\label{eq:stressStrainHB}
\partial_{\gamma}\Sigma&= -1 + v = -1 - D^2 \partial_x^2P(0, \gamma) \quad ,
\end{align}
where we evaluated Eq.\ \eqref{eq:HBquasistatic} at $x=0$ to find $v= - D \partial^2_xP(0, \gamma)/\partial_xP(0, \gamma)$.
Using Eqs.\ (\ref{eq:HBquasistatic}) and (\ref{eq:stressStrainHB}),  $P(x)$ can be computed  for any given  $P_0(x)$, allowing us to compute $\Sigma(\gamma)$ from Eq.\ \eqref{eq:stressStrainHB}.

We demonstrate the existence of a ductile to brittle transition using initial stability distribution $P_0(x)\sim(1-\alpha)\exp{(-(x-1)^2/(2 s_P^2))} + \alpha x (2 - x)$, where $s_P= 0.05$ is kept constant and the distribution is normalized to $1$ on the interval $x \in [0,2]$, as shown in Fig. \ref{fig:meanField}a.
For $\alpha=0.4$  the  stress does not overshoot while for $\alpha= 0.02$ the system is brittle. At an intermediate value $\alpha= 0.08$ the system is still ductile but the stress overshoots, as shown in Fig.\ \ref{fig:meanField}b. Since $P_0(x)$ changes smoothly with $\alpha$ there has to be an $\alpha_c$ at which brittle failure occurs. This transition is continuous and of the usual saddle-node type, so that the magnitude of the stress jump scales as $\Delta \Sigma\sim (\alpha_c-\alpha)^{1/2}$. The same exponents are found in mean field disordered magnets \cite{Ozawa18,Nandi16}. However,  avalanches behave differently than in magnets: from Eq.\ \eqref{eq:averageAvalanche} and the smoothness of the $\Sigma(\epsilon)$ curve, we get for brittle materials that $\langle S\rangle \sim \sqrt{N}/\sqrt{\epsilon_c-\epsilon}$ where $\epsilon_c$ is the strain at which failure occurs. Avalanches statistics can thus be used to forecast $\epsilon_c$.

Our results have an interesting microscopic interpretation in terms of avalanches: from Eqs.\eqref{eq:averageAvalanche} and \eqref{eq:stressStrainHB} we obtain that $\langle S\rangle\sim -\sqrt{N}/ \partial_x^2P(0)$ for $\partial_x^2P(0) < 0$ and it diverges otherwise. The avalanche size is thus controlled by the curvature of $P(x)$ at $x=0$, whereby brittle failure occurs when this curvature vanishes. This result can be rationalized by a simple scaling argument following ideas from \cite{Jagla15}. When an avalanche is initiated, the instantaneous number of unstable elements $n_u$ evolves at each plastic event, and the avalanche ends when $n_u$ returns to $0$. If $P(x)= x/D$, during each plastic event one element is stabilized and on average one element becomes unstable. Therefore, $n_u$ performs a simple random walk and there is no cutoff $S_c$ in the avalanche size distribution. However, if the quadratic term is finite $P(x)= x/D + \partial_x^2P(0) x^2/2$ and a drift appears in the evolution of $n_u$. When $\partial_x^2P(0) < 0$, on average less than one element becomes unstable per plastic event and the drift is negative. $S_c$ corresponds to the avalanche size where the integrated drift $ -N \int_0^{x_c} \partial_x^2P(0) x^2 dx$ is of the order of fluctuations $S_c^{1/2}$. Here, $x_c \sim \sqrt{2 D S_c/N}$ is the characteristic value of the initial stability of elements that became unstable in the avalanche. We thus obtain $S_c \sim -N^{1/2}/\partial_x^{2}P(0)$: the negative curvature of $P(x)$ at $x=0$ determines the avalanche size by depleting the pool of elements that can become unstable.

\begin{figure}[ht!]
\centering
\includegraphics[width=\textwidth]{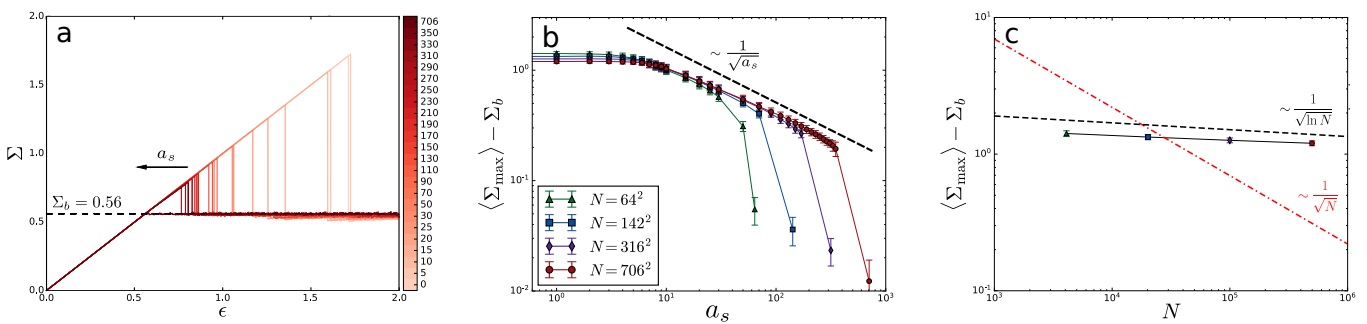}
\caption{a) Stress {\it vs} strain curve in an elasto-plastic model with $N= 706^2$ in which a scar of varying size $a_s$ (as indicated in color) was inserted. b) Maximal stress reached $\Sigma_\mathrm{max}$ as a function of the scar length $a_s$. c) When no scars are inserted, $\Sigma_{\mathrm{max}}$ decreases very slowly with $N$, consistent with our prediction $\Sigma_{\mathrm{max}} - \Sigma_b \sim 1/\sqrt{\ln{N}}$.}
\label{fig:Fig3}
\end{figure}

\paragraph*{Nucleation of shear band}
We now argue that for very brittle materials at least, macroscopic failure can occur without the apparent divergence of avalanche size described by Eq.\ \eqref{eq:averageAvalanche}; and thus cannot be easily anticipated by a growing crackling noise. Instead, a shear band can nucleate in a region which, by chance, is weaker than the rest of the material. Consider a region of dimension $d-1$, where $d$ is the spatial dimension, and of linear extension $a$ that has already yielded, and thus has smaller yield stresses than the rest of the material. We denote by $\Sigma_b$ the shear stress such a narrow shear band can sustain in the limit of large $a$ ($\Sigma_b$ can in general depend on system preparation). If $\Sigma>\Sigma_b$, the stress will be distorted by this weak region. This is a classical calculation of fracture mechanics \cite{Anderson05}, leading to a stress at a distance $r$ to the tip of the shear band of order $\Sigma(r)\sim (\Sigma-\Sigma_b)\sqrt{a}/\sqrt{r}\equiv {\cal K}/\sqrt{r}$ where ${\cal K}$ is called the intensity factor. In analogy with fracture mechanics, we expect the shear band to propagate if ${\cal K}$ is larger than some critical value ${\cal K}_c$, leading to a critical nucleus size $a_c$ triggering failure:
\begin{align}
\label{11}
a_c\sim \frac{1}{(\Sigma-\Sigma_b)^2}
\end{align}
Eq.\ \eqref{11} is easily tested in elasto-plastic models by inserting a ``scar'', i.e.\ a region with unusually small yield stresses of extension $a_s$. This procedure is analogous to the introduction of a void in a material, as is often used to measure its fracture toughness \cite{Anderson05}. From Eq.\ \eqref{11} we expect a brittle failure to occur for some $\Sigma_{\mathrm{max}}$ satisfying $\Sigma_{\mathrm{max}}-\Sigma_b \sim 1/\sqrt{a_s}$. This prediction is confirmed in Fig.\ \ref{fig:Fig3}a,b.

\onecolumngrid

\twocolumngrid

In a large, homogeneously prepared system, spontaneous shear bands will occur. The probability to find a weak region of spatial extension $a$ follows $p(a) \sim N\exp(-a^{d-1})$, the largest weak region formed by chance follows $ a \sim( \ln{N})^{1/(d-1)}$. Together with Eq.\ \eqref{11} this leads to $\Sigma_{\mathrm{max}} - \Sigma_b \sim 1/(\ln{N})^{1/(2(d-1))}$. This decay is so weak that even for $N$ of the order of the Avogadro number, we expect the overshoot to be significant. It is hard to test this asymptotic result numerically. However we find that for the elasto-plastic model, the dependence of $\Sigma_{\mathrm{max}}$ with $N$ is consistent with the slow decay predicted, as shown in Fig.\ \ref{fig:Fig3}c. The data exclude the more rapid decay $1/\sqrt{N}$ expected from a naive central limit theorem argument.

\paragraph*{Conclusion}
Brittleness is one of the most practically important properties of materials. We have shown that elasto-plastic models can reproduce the ductile to brittle transition in amorphous solids
as their initial stability is increased, in agreement with experimental and recent numerical observations. We have explained this result in a mean field approximation, in which macroscopic failure
can always be predicted by a growing crackling noise. We have argued, however, that for very brittle materials, failure is induced by rare events in which a shear band nucleates, which cannot be forecasted,
and we have provided a theoretical description of this nucleation.

Our work suggests interesting venues for further theoretical and experimental studies. Both the anatomy of the shear bands as well as the possibility that failure can be anticipated by crackling noise in some regimes could be investigated systematically in terms of relevant parameters, including system preparation, loading apparatus but also strain rate and temperature, which has recently been incorporated in elasto-plastic models \cite{Liu18}.

\begin{acknowledgements}
We thank L.~Berthier, G.~Biroli, M. Ozawa, G. Tarjus and A. Rosso for sharing unpublished results and discussions and the Simons collaboration for discussions. M.~W.~thanks the Swiss National Science Foundation for support under Grant No. 200021-165509 and the Simons Foundation Grant ($\#$454953 Matthieu Wyart).
T.~G.~was partly financially supported by The Netherlands Organisation for Scientific Research (NWO) by a NWO Rubicon grant number 680-50-1520.
We acknowledge open-source software: the SciPy ecosystem \cite{SciPy} and GNU parallel \cite{Tange11}.
\end{acknowledgements}






\bibliographystyle{unsrtnat}
\bibliography{tom}



\end{document}


\title{Supplementary Material of ``Elasto-plastic description of brittle failure in amorphous materials''}


\maketitle

\section{Implementation of the elasto-plastic model}
We implement a two-dimensional elasto-plastic model on a periodic lattice of sizes $L= 64, 142, 316, 706$. The propagator $G(r, \phi)$ is a periodic version of an infinite system propagator $G_0(r, \phi) \sim \cos{4\phi}/r^2$ and it is normalized so that $G(\vec{r}= 0) = -1$. This propagator preserves the sum of stresses along each row and column of elements. Thus, to keep the sum of stresses in all rows and columns the same during the initialization of the stress distribution $P(\sigma)$ we proceed as follows. We start with $0$ stress in each element. Then, for each element $i$ we draw a random stress $\delta\sigma$ from a normal distribution $\mathcal{N}(0, s_0^2)$ and we draw two random integer numbers $\delta x$ and $\delta y$ between $1$ and the system's length $L$. Then we add the stress $\delta \sigma$ to element $i$ and the element at coordinates $(x_i - \delta x, y_i - \delta y)$ and we subtract $\delta\sigma$ from the stresses of elements at positions $(x_i - \delta x, y_i)$ and $(x_i, y_i - \delta y)$. Periodicity is imposed when needed. Finally, since on average each element has received a stress update $4$ times by a random number drawn from a normal distribution of variance $s_0^2$, we divide the stress of all elements by $2$ to keep the variance of the initial stress distribution equal to $s_0^2$. We use $s_0= 0.45$ in Fig.\ 1 and $s_0= 0.3$ in Fig.\ 4.

The initial distribution of yield stresses $P(\sigma^Y)$ is a normal distribution $\mathcal{N}(m, 0.01)$. In cases 1, 2 and 3 in Fig.\ 1 we use $m= 1.3, 1.5, 1.8$. In Fig.\ 4 $m= 3.0$ except in the ``scar'' region where $m= 1.0$.

After a plastic failure, the yield stress of the element is updated with a random number from a normal distribution $\mathcal{N}(1, 0.01)$, and the stress of the element is set to a random value drawn from a normal distribution $\mathcal{N}(0, 0.01)$.

\bibliographystyle{unsrt}
\bibliography{../../../bib/Wyartbibnew.bib}

